\documentclass[multphys,vecphys]{svmult}

\usepackage{makeidx}         
\usepackage{graphicx}        
\usepackage{multicol}        
\usepackage[bottom]{footmisc}
\usepackage{url}

\makeindex             

\begin{document}

\title*{Multi-wavelength interferometry of evolved stars using VLTI and VLBA}
\author{M.~Wittkowski\inst{1}\and
D.~A.~Boboltz\inst{2}\and
T.~Driebe\inst{3}\and
K.~Ohnaka\inst{3}}
\institute{European Southern Observatory, Garching, Germany
\texttt{mwittkow@eso.org}
\and U.S. Naval Observatory, Washington, DC, USA 
\texttt{dboboltz@usno.navy.mil}
\and Max-Planck-Institut f\"ur Radioastronomie, Bonn, 
Germany \texttt{driebe@mpifr-bonn.mpg.de,kohnaka@mpifr-bonn.mpg.de}}
%
%
\maketitle
\abstract{We report on our project of coordinated VLTI/VLBA 
observations of the atmospheres and circumstellar environments of
evolved stars. We illustrate in general the potential of interferometric 
measurements to study stellar atmospheres and envelopes, and demonstrate 
in particular the advantages of a coordinated multi-wavelength approach 
including near/mid-infrared as well as radio interferometry. 
We have so far made use of VLTI observations of the near- 
and mid-infrared stellar sizes and of concurrent VLBA observations of 
the SiO maser 
emission. To date, this project includes studies of the Mira
stars S\,Ori and RR\,Aql as well as of the supergiant AH\,Sco. 
These sources all show strong silicate emission features in their 
mid-infrared spectra. 
In addition, they each have relatively strong SiO maser emission.  
The results from our first epochs of S\,Ori measurements have recently 
been published \cite{mwi:boboltz05} and the main results 
are reviewed here. The S\,Ori maser ring is found to lie at a mean distance
of about 2 stellar radii, a result that is virtually free of the usual
uncertainty inherent in comparing observations of variable stars 
widely separated in time and stellar phase. We discuss the status of our 
more recent S\,Ori, RR\,Aql, and AH\,Sco observations, and present an 
outlook on the continuation of our project.}
\section{Introduction}
The evolution of cool luminous stars,
including Mira variables, is accompanied by significant mass-loss to the
circumstellar environment (CSE) with mass-loss rates of up
to $10^{-7} - 10^{-4}$\,M$_\odot$/year (e.g. \cite{mwi:jura90}).
The detailed structure of the CSE, the detailed physical nature of the
mass-loss process from evolved stars, and especially
its connection with the pulsation mechanism in the case of Mira
variable stars, are not well understood.
Furthermore, one of the basic unknowns in the study of late-type stars
is the mechanism by which usually spherically symmetric stars on the 
asymptotic giant branch (AGB) evolve to form axisymmetric or bipolar
planetary nebulae (PNe). Possible origins of asymmetric structures
include, among others, binarity, capture of substellar companions,
stellar rotation, or 
magnetic fields. While it is generally believed that the observed
pronounced asymmetries of the envelopes of PNe form when the star 
evolves from the tip of the AGB branch toward the blue part of the HR diagram,
there is evidence for some 
asymmetric structures already around AGB stars and supergiants 
(e.g., \cite{mwi:weigelt96,mwi:weigelt98,mwi:wittkowski98,mwi:monnier99,
mwi:boboltz00}).

\begin{figure}[tb]
\centering
\includegraphics[width=0.98\textwidth]{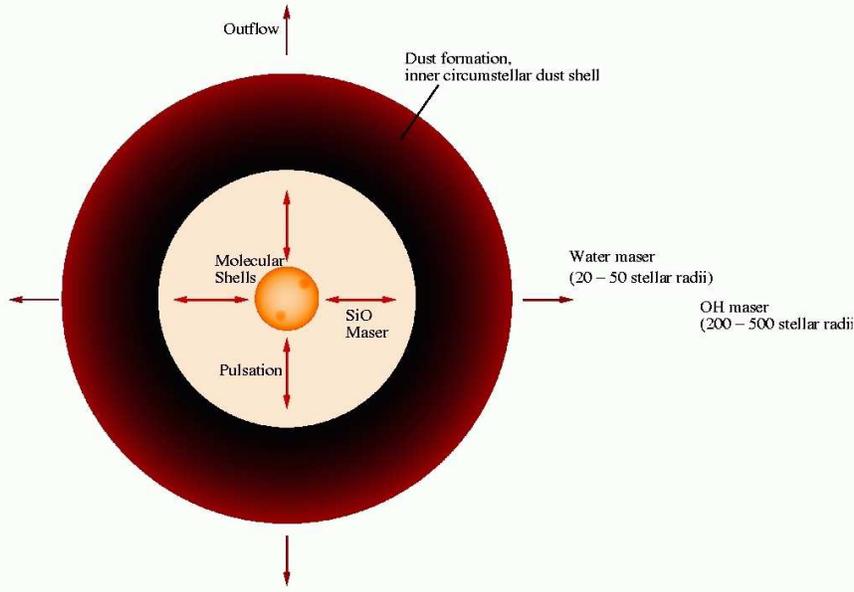}
\caption{Sketch of a Mira variable star and its circumstellar
envelope (CSE). A multi-wavelength study (MIDI/AMBER/VLBA) is well suited
to probe the different regions shown here. Owing to the stellar variability,
only contemporaneous observations are meaningful.}
\label{mwi:scheme}
\end{figure}
\begin{figure}[tb]
\centering
\includegraphics[width=0.98\textwidth]{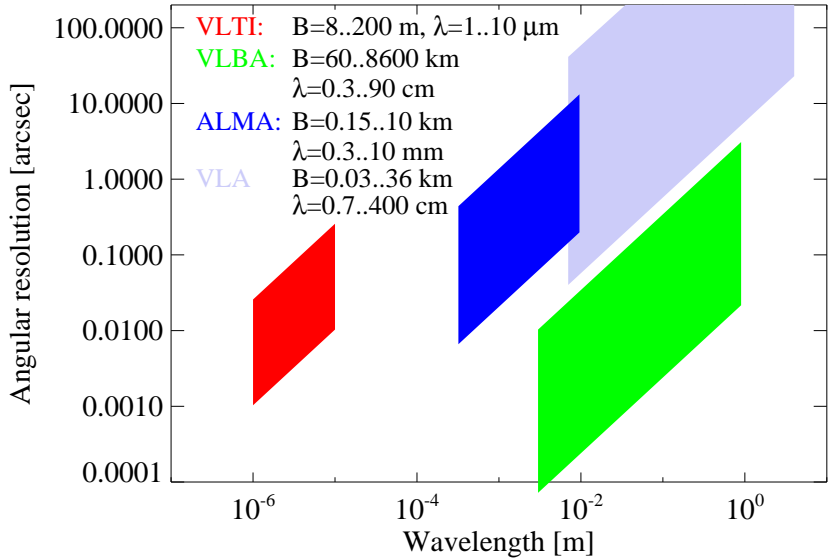}
\caption{Comparison of resolution and wavelength ranges
of the infrared, millimeter, and radio interferometric facilities
VLTI, ALMA, VLBA, and VLA.}
\label{mwi:facilities}
\end{figure}
Coordinated multi-wavelength studies (near-infrared, mid-infrared,
radio, millimeter) of the stellar surface (photosphere) {\it and} the CSE 
at different
distances from the stellar photosphere and obtained at corresponding 
cycle/phase values of the stellar variability curve are best suited to
improve our general understanding of the atmospheric structure, the CSE, 
the mass-loss process, and ultimately of the evolution of 
symmetric AGB stars toward axisymmetric or bipolar planetary nebulae. 
Fig.~\ref{mwi:scheme} shows a schematic view of a Mira variable star, 
indicating the different regions that can be probed by different 
techniques/wavelength ranges (VLTI/AMBER, VLTI/MIDI, VLBA/maser, ALMA).
Fig.~\ref{mwi:facilities} shows a comparison of the VLTI, VLBA, ALMA, and
VLA interferometric facilities in terms of wavelength ranges and angular
resolution. VLTI, VLBA, and ALMA allow us to observe the same
evolved stars in terms of sensitivity, and reach a comparable angular
resolution at their respective wavelength ranges.

The conditions near the stellar surface can best be studied
by means of optical/near-infrared long-baseline interferometry.
This technique has provided information regarding the stellar photospheric
diameter, asymmetries/surface inhomogeneities, effective temperature, and
center-to-limb intensity variations including the effects of close
molecular shells, for a number of non-Mira and Mira giants
(see, e.g., \cite{mwi:haniff95,mwi:wittkowski01,mwi:thompson02,mwi:hofmann02,
mwi:wittkowski04,mwi:woodruff04,mwi:boboltz05,mwi:fedele05}).

The structure and physical parameters of the molecular shells located
between the photosphere and the dust formation zone, as well as of the
dust shell itself can be probed by mid-infrared interferometry 
(e.g. \cite{mwi:danchi94}).
This has also recently been demonstrated by using the 
spectro-interferometric capabilities of the VLTI/MIDI instrument 
to study the Mira star\,RR Sco \cite{mwi:ohnaka05}. The
model obtained in this work includes a warm molecular (SiO and H$_2$O)
layer as well as a dust shell of corundum and silicate, and can well
reproduce the obtained MIDI visibility values.

Complementary information regarding the molecular shells can be
obtained by observing the maser radiation that some of these molecules
emit. The structure and dynamics of the CSE of Mira variables and other
evolved stars has been investigated by mapping SiO maser emission
at typically about 2 stellar radii toward these stars using
very long baseline interferometry (VLBI) at radio wavelengths
(e.g., \cite{mwi:boboltz97,mwi:kemball97,mwi:boboltz05}).

Results regarding the relationships between the different regions
mentioned above and shown in Fig.~\ref{mwi:scheme} suffer often from 
uncertainties inherent in comparing observations of variable stars widely 
separated in time and stellar phase 
(see the discussion in \cite{mwi:boboltz05}). Both, the
photospheric stellar size as well as the mean diameter of the
SiO maser shell are known to vary as a function of the 
stellar variability phase with amplitudes of 20-50\% (see \cite{mwi:ireland04}
for theoretical and \cite{mwi:thompson02} for observational estimates
of the variability of the stellar diameter; as well as \cite{mwi:humphreys02}
for theoretical and \cite{mwi:diamond03} for observational estimates
of the variability of the mean SiO maser ring diameter).
  
To overcome these limitations, we have established
a program of coordinated and concurrent observations at near-infrared, 
mid-infrared, and radio wavelengths of evolved stars, aiming at a better 
understanding of the structure of the CSE, of the mass-loss process, and of
the triggering and formation of asymmetric structures. 

In the following, we describe recent results obtained with optical/infrared
interferometry on the stellar atmospheric structure
of regular non-Mira (Section 2) and Mira (Section 3) giants.
Our joint VLTI/VLBA observations of the Mira star S\,Ori are discussed
in Section 4. Finally, we give an outlook (Section 5) on further 
measurements and future ideas. The latter includes desirable 
2nd generation instrumentation based on the requirements of this
particular project alone.
\section{The atmospheric structure of non-Mira giants}
Fundamental parameters, most importantly radii and effective temperatures,
of regular cool giant stars have frequently been obtained
with interferometric and other high angular resolution techniques,
thanks to the favorable brightness and size of these stars.
Further parameters of the stellar structure, as the strength
of the limb-darkening effect, can be studied when more than one resolution
element across the stellar disk is employed.
Through the direct measurement of the center-to-limb intensity 
variation (CLV) across stellar disks and their close environments, 
interferometry probes the vertical temperature profile, as well as 
horizontal inhomogeneities. However, the required direct measurements 
of stellar intensity profiles are among the most challenging programs 
in modern optical interferometry. Since more than one resolution element 
across the stellar disk is needed to determine surface structure parameters 
beyond diameters, the long baselines needed to
obtain this resolution also produce very low visibility amplitudes
corresponding to vanishing fringe contrasts.
Such direct limb-darkening studies have been accomplished for
a relatively small number of stars using different interferometric
facilities (including, for instance, \cite{mwi:hanbury73,mwi:quirrenbach96,
mwi:hajian98,mwi:wittkowski01,mwi:wittkowski04}).

\begin{figure}[tb]
\centering
\includegraphics[width=0.32\textwidth]{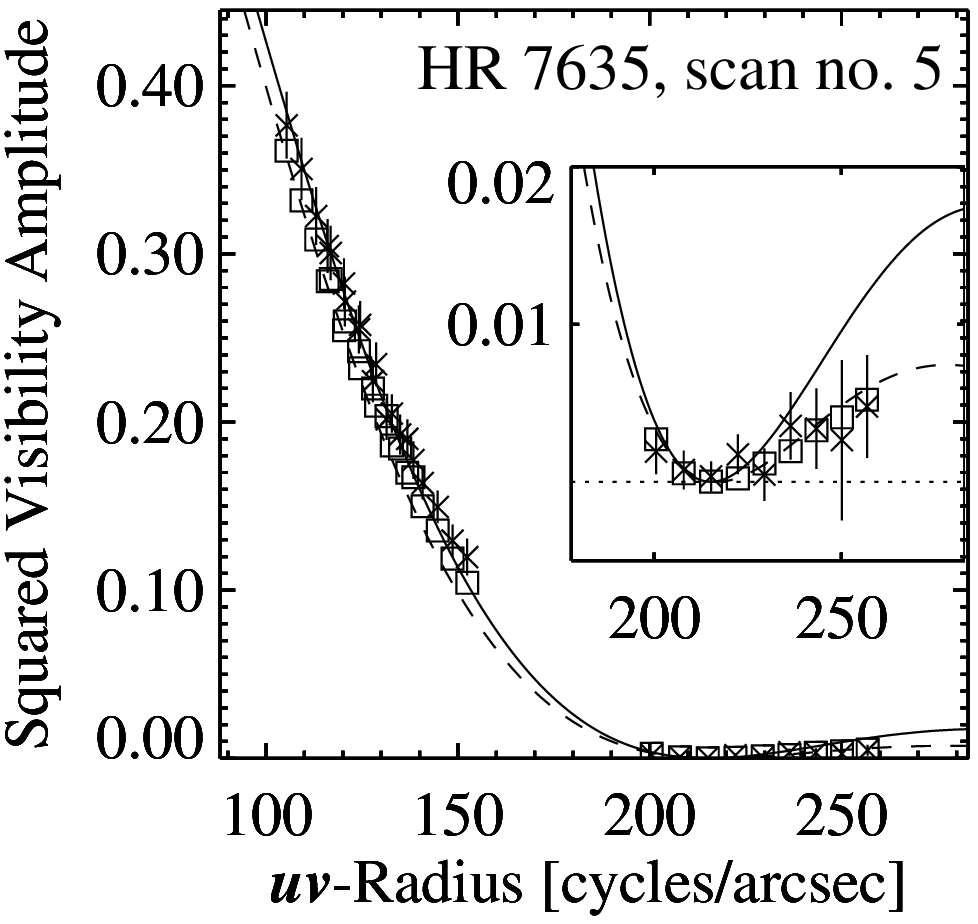}
\includegraphics[width=0.32\textwidth]{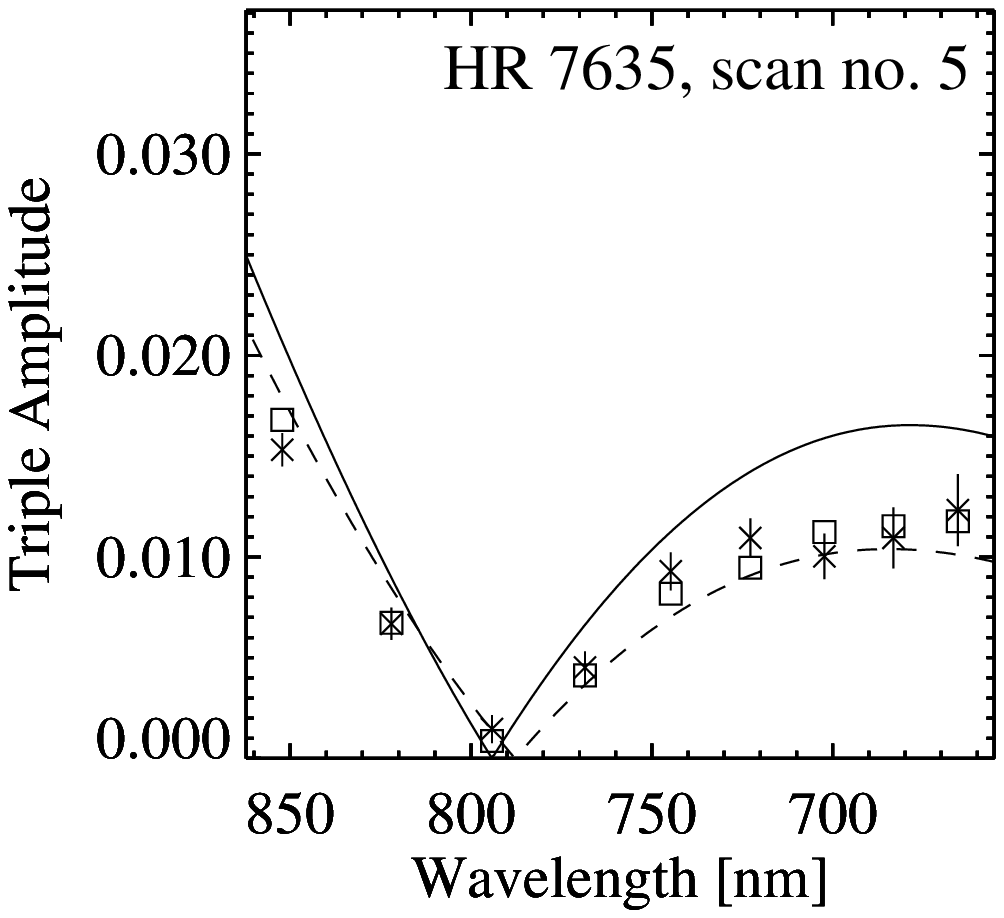}
\includegraphics[width=0.32\textwidth]{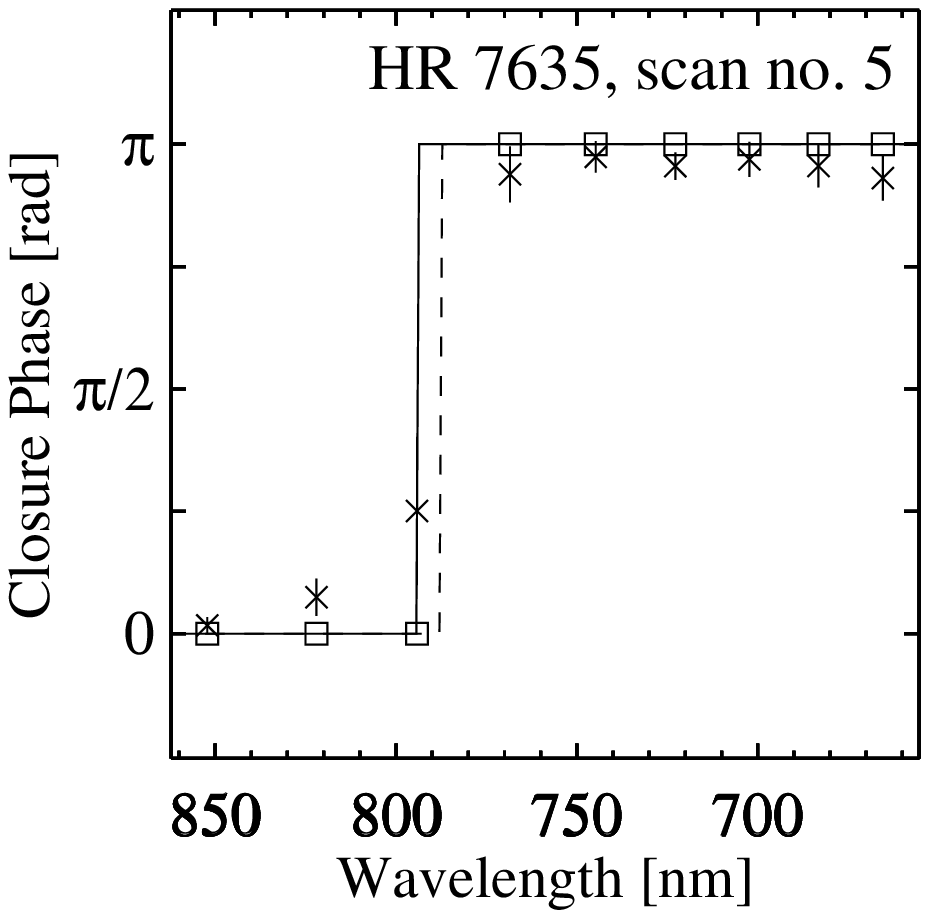}
\caption{NPOI limb-darkening observations (squared visibility amplitude,
triple amplitude, closure phase) of the M0 giant $\gamma$\,Sge, together
with a comparison to the best fitting {\tt ATLAS\,9} model atmosphere 
prediction (squares).
For comparison, the solid line denotes a uniform disk model, and the
dashed line a fully-darkened disk model. {\tt ATLAS\,9} models with
variations of $T_\mathrm{eff}$ and $\log g$ result in significantly different
model predictions. From \cite{mwi:wittkowski01}.}
\label{mwi:gamsge}
\end{figure}
Recent optical multi-wavelength measurements
of the cool giants $\gamma$\,Sge and BY\,Boo
\cite{mwi:wittkowski01} succeeded not only in
directly detecting the limb-darkening
effect, but also in constraining {\tt ATLAS\,9} (\cite{mwi:kurucz93})
model atmosphere parameters. Fig.~\ref{mwi:gamsge} shows
one dataset including squared visibility amplitudes, triple amplitudes,
and closure phases of the M0 giant $\gamma$\,Sge obtained with NPOI,
together with a comparison to the best fitting {\tt ATLAS\,9} model atmosphere 
prediction. {\tt ATLAS\,9} models with
variations of $T_\mathrm{eff}$ and $\log g$ result in significantly different
model predictions. By this direct comparison of the NPOI data to the
{\tt ATLAS 9} models alone, the effective temperature of $\gamma$\,Sge
is constrained to 4160$\pm$100\,K. The limb-darkening observations
are less sensitive to variations of the surface gravity, and $\log g$
is constrained to 0.9$\pm$1.0 \cite{mwi:wittkowski01}. These constraints
are well consistent with independent estimates, such as calibrations
of the spectral type. 
Furthermore, it was shown that these interferometric and
spectroscopic measurements of $\gamma$\,Sge both compare well 
with predictions by the same spherical
{\tt PHOENIX} \cite{mwi:hauschildt99} model atmosphere 
\cite{mwi:aufdenberg03}.

\begin{figure}[tb]
\centering
\includegraphics[width=0.6\textwidth]{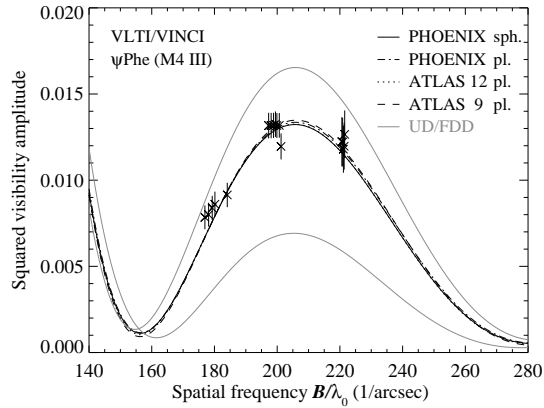}
\caption{VLTI limb-darkening observations of the M4 giant $\psi$\,Phe
\cite{mwi:wittkowski04}.}
\label{mwi:psiphe}
\end{figure}
The first
limb-darkening observation that was obtained with the VLTI succeeded
in the early commissioning phase of the VLTI \cite{mwi:wittkowski04}.
Using the VINCI instrument, $K$-band visibilities of the
M4 giant $\psi$\,Phe were measured in the first and second lobe of the 
visibility function.
These observations were found to be consistent with predictions
by {\tt PHOENIX} and {\tt ATLAS} model atmospheres, the parameters for
which were constrained by comparison to available spectrophotometry and
theoretical stellar evolutionary tracks (see Fig.~\ref{mwi:psiphe}). 
Such limb-darkening
observations also result in very precise and accurate
radius estimates because of the precise description of the CLV. Future use
of the spectro-interferometric capabilities of AMBER and MIDI will
enable us to study the wavelength-dependence of the limb-darkening effect,
which results in stronger tests and constraints of the model atmospheres
than these broad-band observations (cf. the wavelength-dependent optical
studies with NPOI as described above).

Another strong test of model atmospheres is the direct comparison
of spectro-photometry, high-resolution spectra, and limb-darkening
observations to predictions by the same model atmosphere. Such studies
are presented in these proceedings (V.~Roccatagliata et al.). 
Available spectrophotometry, high-resolution UVES ultraviolet/optical
spectra, as well as near-infrared VLTI/VINCI $K$-band limb-darkening 
measurements are compared to predictions by {\tt PHOENIX} model atmospheres,
and good agreement is found (see Figs. 1 and 2 in Roccatagliata et al., these
proceedings).
\section{The atmospheric structure of Mira giants}
For cool pulsating Mira stars, the CLVs are expected to be more complex
than for non-pulsating M giants due to the effects of molecular layers
close to the continuum-forming layers.
Based on self-excited hydrodynamic model
atmospheres of Mira stars (\cite{mwi:hofmann98,
mwi:tej03,mwi:ireland04}, Scholz \& Wood, private
communication), broad-band CLVs may indeed appear 
as Gaussian-shaped or multi-component functions, and to exhibit
temporal variations as a 
function of stellar phase and cycle, in accordance with observations
(see introduction). These complex shapes of the CLV make it difficult to 
define an appropriate stellar radius. Different radius definitions, such
as the Rosseland mean radius, the continuum radius, or the radius at which 
the filter-averaged intensity drops by 50\%, may result in different
values for the same CLV. For complex CLVs at certain variability phases 
these definitions can result in differences of up to about 20\%
(on these topics, see also \cite{mwi:scholz03}).
However, interferometric measurements covering a sufficiently
wide range of spatial frequencies
can directly be compared to CLV predictions by model atmospheres
without the need of a particular radius definition. At pre-maximum
stellar phases, when the temperature is highest, the broad-band CLVs are 
less contaminated by molecular layers, and different radius definitions
agree relatively well (Scholz \& Wood, private communication).

$K$-band VINCI observations of the prototype
Mira stars  $o$\,Cet and R\,Leo have been presented by \cite{mwi:woodruff04}
and \cite{mwi:fedele05}, respectively. These measurements are also
desribed in more detail elsewhere in these 
proceedings (Driebe et al., Fedele et al.). 
These measurements at post-maximum stellar phases
indicate indeed $K$-band CLVs which are clearly different from a
uniform disk profile already in the first lobe of the visibility
function. The measured visibility values were found to
be consistent with predictions by the self-excited dynamic
Mira model atmospheres described above that include molecular
shells close to continuum-forming layers.
\section{Joint VLTI/VLBA observations of the Mira star S\,Ori}
\begin{figure}[tb]
\centering
\includegraphics[width=0.65\textwidth]{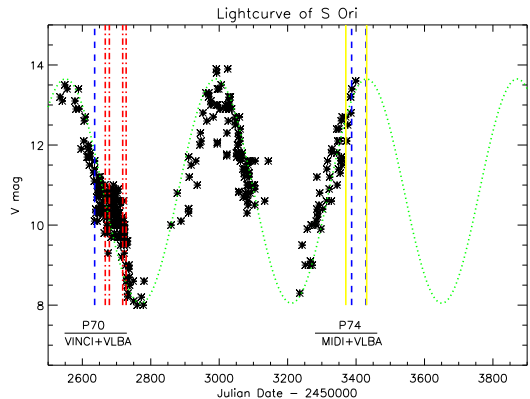}
\caption{Lightcurve of S\,Ori together with the epochs of
our joint VLTI/VLBA measurements obtained so far. Note that the y-axis 
is given with increasing $V$ magnitude, i.e. the stellar maximum is at the 
bottom and stellar minimum at the top. The study of S\,Ori was started
in ESO period P70 (Dec. 2002/Jan. 2003) including near-infrared
$K$-band VINCI and VLBA/SiO maser observations \cite{mwi:boboltz05}.
In December 2004/January 2005, we obtained concurrent mid-infrared 
VLTI/MIDI and VLBA/SiO maser observations.}
\label{mwi:lightcurve}
\end{figure}
\begin{figure}[tb]
\centering
\includegraphics[width=0.49\textwidth]{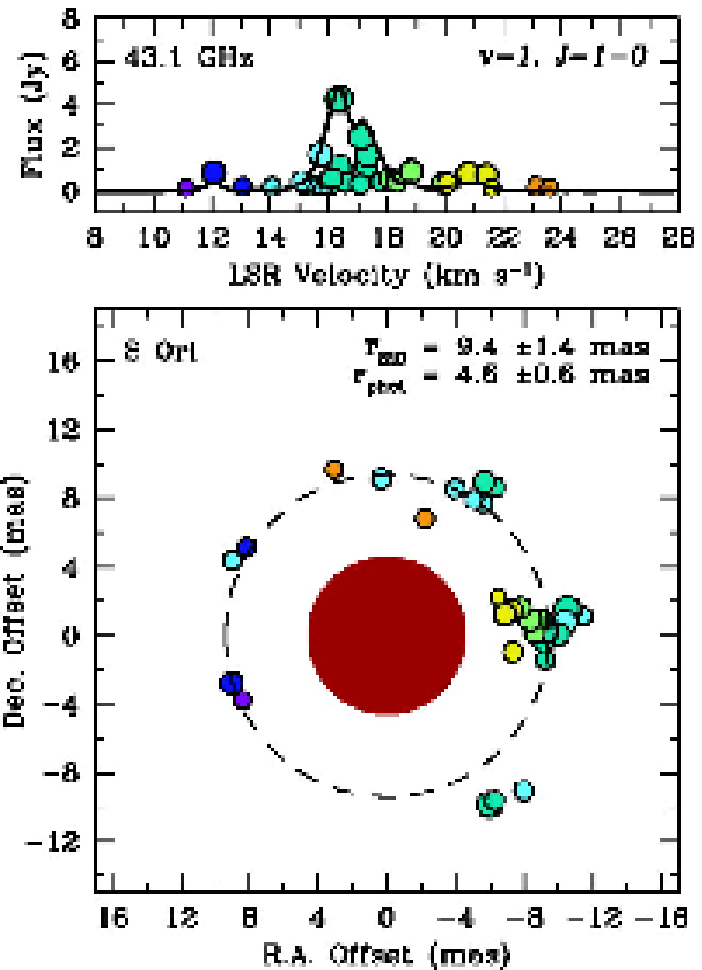}
\includegraphics[width=0.49\textwidth]{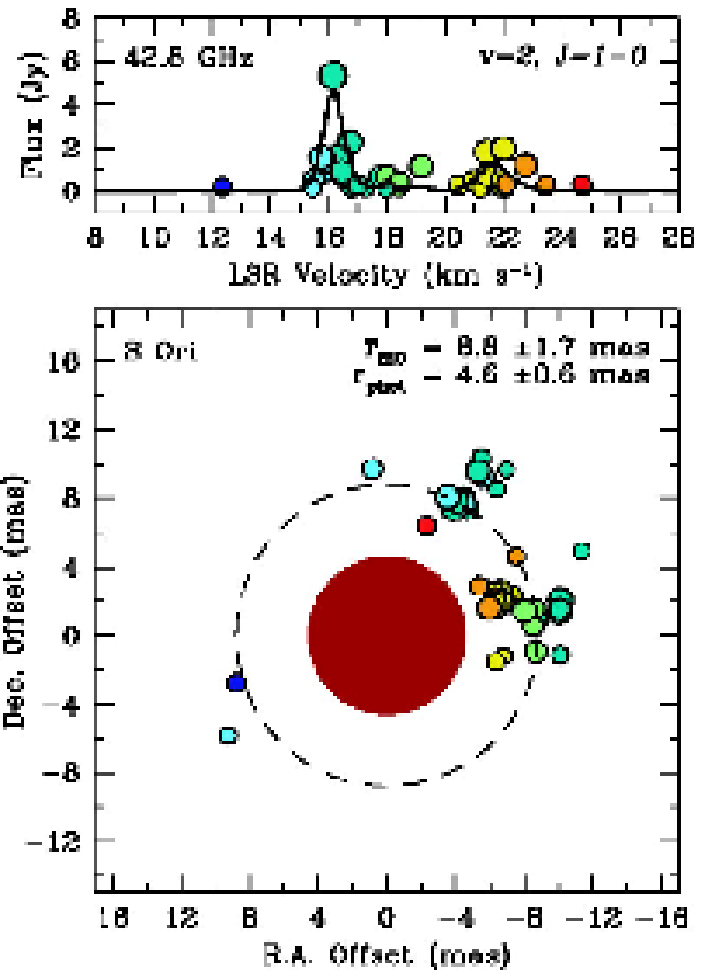}
\caption{First-ever coordinated observations between ESO's VLTI and
NRAO's VLBA facilities: SiO maser emissions toward the Mira variable
S\,Ori measured with the VLBA,
together with the near-infrared diameter measured quasi simultaneously
with the VLTI (red stellar disk). The left panels shows the 43.1\,GHz
maser transition, and the right panel the 42.8\,GHz transition. 
From \cite{mwi:boboltz05}.}
\label{mwi:sori}
\end{figure}
We started our project of joint VLTI/VLBA observations of Mira stars
in December 2002/January 2003 with coordinated near-infrared 
$K$-band VLTI/VINCI observations of the stellar diameter of the Mira 
variable S\,Ori and quasi-simultaneous VLBA observations of the 43.1\,GHz 
and 42.8\,GHz SiO maser emissions toward this star \cite{mwi:boboltz05}.
We obtained in December 2004/January 2005 further concurrent observations
including mid-infrared VLTI/MIDI observations to probe the molecular
layers and the dust shell of S\,Ori, and new epochs of VLBA observations
of the 43.1\,GHz and 42.8 GHz SiO maser rings.
 
The December 2002/January 2003 observations represent the first-ever 
coordinated observations
between the VLTI and VLBA facilities, and the results from these observations
were recently published \cite{mwi:boboltz05}.
Analysis of the SiO maser data recorded at
a visual variability phase 0.73 show the average distance of the masers
from the center of the distribution to be 9.4~mas for the 
$v=1, J=1-0$ (43.1 GHz) masers and 8.8~mas for the $v=2, J=1-0$ (42.8 GHz) 
masers. The velocity structure of the
SiO masers appears to be random with no significant indication of
global expansion/infall or rotation.
The determined near-infrared, $K$-band, uniform disk (UD) diameters
decreased from $\sim$\,10.5\,mas at phase 0.80 to $\sim$10.2\,mas at
phase 0.95.  For the epoch of our VLBA measurements,
an extrapolated UD diameter of $\Theta_\mathrm{UD}^K=10.8 \pm 0.3$\,mas
was obtained, corresponding to a linear radius
of $R_\mathrm{UD}^K = 2.3 \pm 0.5$~AU or
$R_\mathrm{UD}^K =490 \pm 115~R_\odot$. The model predicted difference 
between the continuum and $K$-band UD diameters is relatively low in the 
pre-maximum region of the visual variability curve as in the case of our 
observations (see above). At this phase of 0.73, the continuum diameter may 
be smaller than the $K$-band UD diameter by about 15\% \cite{mwi:ireland04}.
With this assumption, the continuum photospheric diameter
for the epoch of our VLBA observation would be
$\Theta_\mathrm{Phot}(\mathrm{VLBA\ epoch,\ phase=0.73})\approx 9.2$\,mas.
Our coordinated VLBA/VLTI measurements show that the masers lie
relatively close to the stellar photosphere at a distance of $\sim$\,2
photospheric radii, consistent with model estimates \cite{mwi:humphreys02}
and observations of other Mira stars \cite{mwi:cotton04}.  
This result is virtually free of the usual uncertainty inherent in 
comparing observations of variable stars widely separated in time and 
stellar phase.\\
The new 2004/2005 VLTI and VLBA data are currently being reduced and
analyzed.
\section{Outlook}
We are concentrating on a few stars in order to understand the CSE for a
few sources in depth. In addition to the S\,Ori data described
above, we have to date VLTI/MIDI observations of the supergiant
AH\,Sco (Jul. 2005, Aug. 2005), and of the Mira star 
RR\,Aql (Jul. 2005, Aug. 2005), as well as concurrent VLBA observations for 
each of these targets/epochs. These data are currently being analyzed.
There may be hints toward an inherent difference in the structure
between Mira variables and supergiants, in particular regarding the 
relative distances of photosphere, SiO maser ring, and inner dust shell
boundary (cf. \cite{mwi:danchi94,mwi:boboltz05}).

Further studies will aim at including more detailed near-infrared studies
of the stellar atmospheric structure (close to the photosphere) employing
VLTI/AMBER, concurrent with VLTI/MIDI and VLBA observations as
discussed above. Making use of the spectro-interferometric capabilities
of AMBER, and also of the closure-phase information, these studies can
in principle also reveil horizontal surface inhomogeneities (see, e.g.
\cite{mwi:wittkowski02}).

A further step toward our better understanding of the stellar mass-loss
process are interferometric measurements of post-AGB stars.
The first detection of the envelope which surrounds
the post-AGB binary source HR\,4049, by $K$-band VINCI observations,
was recently reported by \cite{mwi:antoniucci05}.
A physical size of the envelope in the near-infrared
$K$-band of about 15 AU (Gaussian FWHM) was derived.
These measurements provide information on the geometry of the emitting
region and cover a range of position angles of about 60 deg. They show
that there is only a slight variation of the size with position angle
covered within this range. These observations are, thus, consistent
with a spherical envelope at this distance from the stellar source,
while an asymmetric envelope cannot be completely ruled out due to the
limitation in azimuth range, spatial frequency, and wavelength range.
Further investigations using
the near-infrared instrument AMBER can reveal the geometry of this
near-infrared component in more detail, and MIDI observations can add
information on cooler dust at larger distances from the stellar surface.

In the more distant future, when second generation instruments at the
VLTI become available, a very valuable addition and 
continuation of this project would be an improved imaging capability
of the VLTI, both at near-infrared as well as at mid-infrared wavelengths.
This would enable us to detect and correlate asymmetric 
structures at the stellar surface and dust shell in a much more precise
way, and hence to better 
understand the transition from spherically symmetric AGB stars to 
axisymmetric or bipolar planetary nebulae. Improved imaging capabilities
can be reached by an increased number of simultaneously combined 
beams. Furthermore, an improved spatial resolution (by using longer
baselines) would be desirable to better match the high angular resolution 
of the VLBA. 
\printindex
\end{document}